\documentclass[achemso,superscriptaddress, twocolumn]{revtex4-2} 
\usepackage{mathtools}
\usepackage{amsmath}
\usepackage{multirow}
\usepackage[english]{babel}
\usepackage{amsmath,lipsum}
\usepackage{etoolbox}
\usepackage{mathrsfs}
\appto{\normalsize}{\zerodisplayskips}
\appto{\small}{\zerodisplayskips}
\appto{\footnotesize}{\zerodisplayskips}
\usepackage{textcomp}
\usepackage{gensymb}
\newcommand{\zerodisplayskips}{%
  \setlength{\abovedisplayskip}{5pt}%
  \setlength{\belowdisplayskip}{5pt}%
  \setlength{\abovedisplayshortskip}{5pt}%
  \setlength{\belowdisplayshortskip}{5pt} }
\appto{\normalsize}{\zerodisplayskips}
\appto{\small}{\zerodisplayskips}
\appto{\footnotesize}{\zerodisplayskips}
\usepackage{graphicx}
\usepackage{color}
\usepackage{enumitem}
\usepackage[dvipsnames]{xcolor}
\usepackage[normalem]{ulem}
\usepackage{hyperref}

\makeatletter
    \def\@seccntformat#1{\@ifundefined{#1@cntformat}%
       {\csname the#1\endcsname\space}%    default
       {\csname #1@cntformat\endcsname}}%  enable individual control
    \def\subsection@cntformat{\thesection.\thesubsection\space} 
    \def\subsubsection@cntformat{\thesection.\thesubsection.\thesubsubsection\space}
\makeatother

\usepackage{fancyhdr}
\usepackage{blindtext}
\DeclareUnicodeCharacter{0301}{\'{e}} % Foreign accent
\DeclareUnicodeCharacter{2009}{\,} % Foreign accent
\DeclareUnicodeCharacter{2212}{-}

\begin{document}

\title{Enhanced sampling of Crystal Nucleation with  Graph Representation Learnt Variables}

\author{Ziyue Zou}
\affiliation{Department of Chemistry and Biochemistry, University of Maryland, College Park 20742, USA.}

\author{Pratyush Tiwary \thanks{ptiwary@umd.edu}}
\email{ptiwary@umd.edu}
\affiliation{Department of Chemistry and Biochemistry, University of Maryland, College Park 20742, USA.}
\affiliation{Institute for Physical Science and Technology, University of Maryland, College Park 20742, USA.}

\date{\today}


\begin{abstract}
\section*{Abstract}
\label{sec:abstract}

In this study, we present a graph neural network-based learning approach using an autoencoder setup to derive low-dimensional variables from features observed in experimental crystal structures. These variables are then biased in enhanced sampling to observe state-to-state transitions and reliable thermodynamic weights. Our approach uses simple convolution and pooling methods. To verify the effectiveness of our protocol, we examined the nucleation of various allotropes and polymorphs of iron and glycine from their molten states. Our graph latent variables when biased in well-tempered metadynamics consistently show transitions between states and achieve accurate free energy calculations in agreement with experiments, both of which are indicators of dependable sampling. This underscores the strength and promise of our graph neural net variables for improved sampling. The protocol shown here should be applicable for other systems and with other sampling methods.

%Here we introduce a representation learning framework to distinguish different polymorphs and allotropes of a given system from one other and perform bias sampling that gives correct relative thermodynamic probabilities of these phases. Our method uses graph neural nets with simple yet powerful convolutional layers to construct low-dimensional variables with an autoencoder architecture that classifies crystal structures based on configurational information from all-atom molecular dynamics. We show a protocol where these   In combination with metadynamics, frequent state-to-state transitions are sampled in two selected representative atomic/molecular systems, namely, iron and zwitterionic glycine which are well-known of their enrichment in allotropy/polymorphism. Thermodynamics properties, free energy differences among metastable states, are computed from the collected samples. The thermodynamic rankings obtained consequentially are compared with literature measurements and show good agreements, which indicates the proposed framework is useful in learning and identifying highly ordered crystal configurations from each other and is able to approximate the reaction coordinates which can be beneficial to many other enhanced sampling methods besides metadynamics.

\end{abstract}

\maketitle
\thispagestyle{fancy}

\section{Introduction}
\label{sec:introduction}

The time-scale problem in the computational study of rare events such as protein folding or crystal nucleation with conventional molecular dynamics (MD) simulations is well-known. For many crystal nucleation processes, the simulation time it takes to witness phase transitions can often range from milliseconds to minutes. However, limited by the vibrational motions of hydrogen bonds, the time step of integration of the equation of motion in typical MD simulations is confined to one or two femtoseconds, which means observing one nucleation event requires years of simulation. Obtaining statistically relevant observations on thermodynamics or kinetics becomes out of the question. Many enhanced sampling methods have been proposed to resolve the problems as mentioned above. \cite{henin2022enhanced} A larger class of such methods belong to the collective variable family, where relevant slow degrees of freedom for the processes of interest are accelerated in a controllable manner. In popular methods such as metadynamics\cite{laio2002escaping}, umbrella sampling\cite{Torrie1977US} or forward flux sampling\cite{PhysRevLett2005FFS}, for practical purposes it is desirable to focus on a maximum of one to three slow degrees of freedom. Ideally these should approximate the reaction coordinate (RC) for the process being studied. \cite{bussi2020using}
In order to design such an approximate RC for the study of rare events, generally one constructs them as a combination of a larger dictionary of features that can collectively distinguish between different metastable states of interest.  To mitigate potential ambiguity, we refer to these features as order parameters (OPs) throughout this work.

Over the years a vast range of such hand-crafted and machine-learnt OPs have been proposed for the study of crystal nucleation. These can be split into different classes. A first class includes task-specific OPs whose definitions rely on particular orientations of particles or molecules and their local environments in the corresponding crystalline packings of interest. \cite{steinhardt1983op,Trout2011orientationop,gavezzotti2011can,Salvalaglio2012uncovering,yi2013molecular,Giberti2013,EnvSim2019Piaggi} A second class of OPs is more generic and does not need prior knowledge of the relevant crystalline packings. These rely on the computation of the exact or approximated thermodynamic observables. Examples are approximate entropy, enthalpy \cite{Piaggi2017entropy,piaggi2018predicting} and moments of coordination number \cite{tsai2019reaction,Finney2022naclnucleation}. Given their generic nature, these OPs can be applied to systems without any prior knowledge for the exploration of the free energy landscape and screening of metastable allotropes or polymorphs. However, it is important to note that this generality may occasionally lead to slow convergence and inefficiency in computing free energy. \cite{giberti2015crystalnucleation}
%\pt{reference for last point?}

% \pt{metadynamics should come later in intro as your method is more generally applicable. emphasize this in conclusion also that while here you show for metad it should work elsewhere. description of what metad is should be in methods section not intro}
%and in this work, we focus on one of such methods, well-tempered metadynamics (WTmetaD) \cite{laio2002escaping,barducci2008well}. In short, systems normally trapped in their initial state are pushed out and escaped by frequent depositions of Gaussian kernels for metadynamics-family methods. In particular, the well-tempered variant allows history-dependent Gaussian kernels being deposited and leads to better convergence of the sampled free energy surface compared to conventional metadynamics. Still, one relatively low-dimensional (typically 1-3D) coordinate needs to be provided for Gaussian depositions in metadynamics under the realization of the curse of dimensionality. Therefore, obtaining the reaction coordinate (RC) becomes one of the important tasks for applications of metadynamics to systems of interest.

\begin{figure*}
  \centering
  \includegraphics[keepaspectratio, width=16cm]{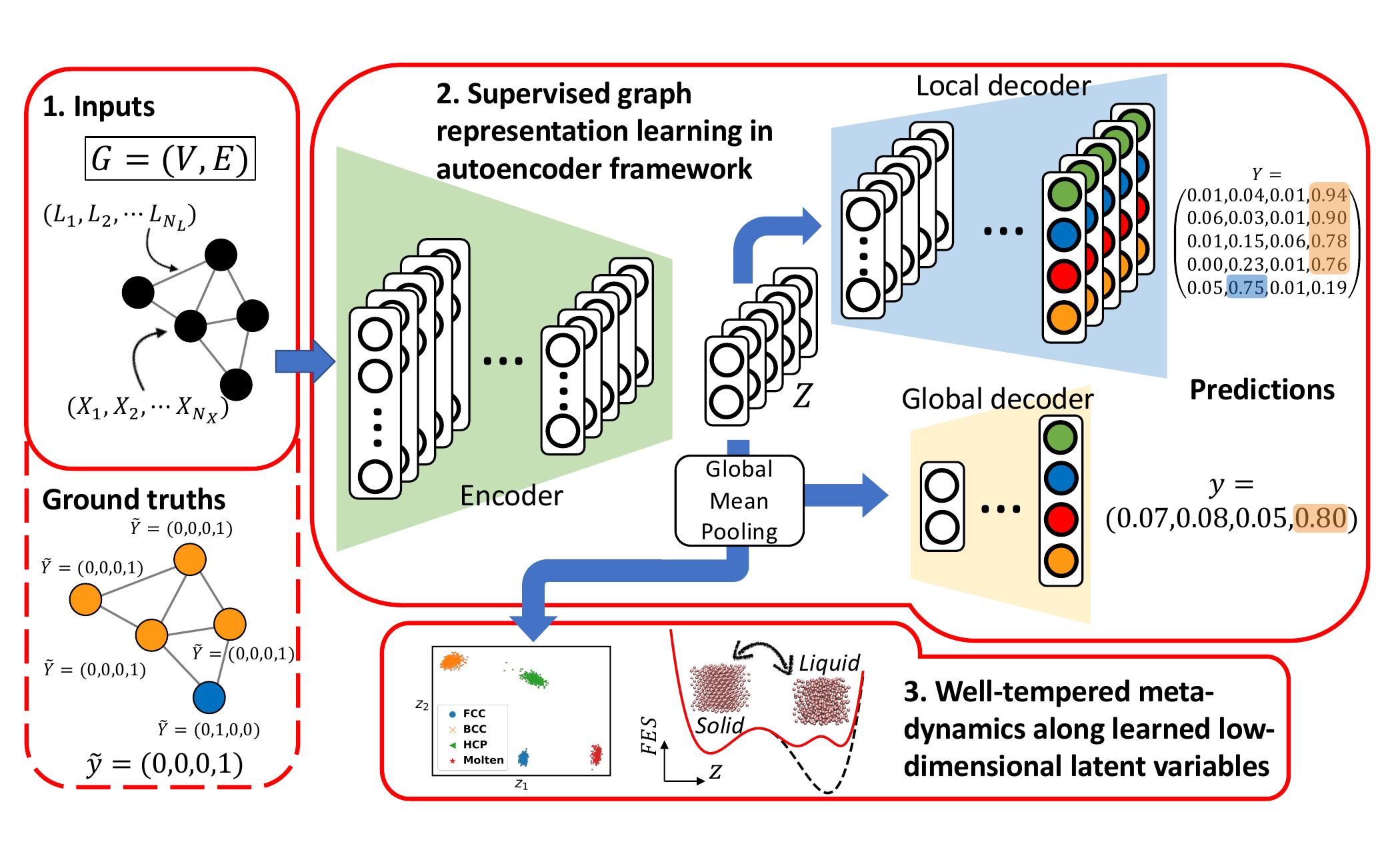}
  \caption
  {Schematic of the pipeline of the graph representation learning architecture. In part 1, atomic coordinates generated from MD simulations are first converted into graphs consist of tensors of node and edge features (X and L respectively; in solid red box) with labels (in dashed box). In part 2, graph data is fed in the GNN model for the training process. Under a supervised learning scheme, ground truths (i.e. labels) are applied for computing and backpropagating the loss. Once the model is trained, the frozen encoder part along with a global mean pooling layer provides latent variables computed on-the-fly as reaction coordinates for metadynamics (part 3).
  }
  \label{fig:gnn}
\end{figure*}

Different from the above two hand-crafted classes of OPs, recent breakthroughs in machine learning (ML) techniques have given rise to a range of neural network (NN) based OPs for a variety of problems, including crystal nucleation. The inherent differentiability of these OPs makes them suited for various enhanced sampling methods that involve the modification of a system's Hamiltonian. \cite{noe2020machine,chen2021collective,sarupria2022machine,beyerle2023recent,mehdi2023enhanced,jung2023machine,dietrich2023machine,herringer2023permutationally,elishav2023collective} We specifically highlight graph neural networks (GNNs) which have emerged as powerful tool in the realm of materials science, including but not limited to efficient descriptions of material energetics \cite{gilmer2017neural,schutt2017schnet}, accurate predictions on material properties \cite{xie2018crystal, jorgensen2018neural}, and robust classifiers of crystal structures and defects \cite{defever2019generalized,moradzadeh2023topology,banik2023cegann,kim2020gcicenet,fulford2019deepice}. 

% \pt{before going into next para on GNNs, you should add one or two sentences about ``Early work involving graph-theoretic ideas, predating the development of GNNs, was already reported for crystal nucleation by ??, ?? and others (talk about gareth, SOAP, fabio pietrucci papers}

Early work involving graph-theoretic ideas, predating the development of GNNs, was already reported for crystal nucleation. Examples include social permutation invariant (SPRINT) coordinates by Pietrucci and Andreoni \cite{pietrucci2011graph}, smooth overlap of atomic positions (SOAP) fingerprints by Bart\'{o}k \textit{et al.} \cite{bartok2013representing}, and other approaches.\cite{Tribello2017analyzing}
%cluster descriptor with depth first search graph reduction algorithm by Tribello and coworkers \cite{Tribello2017analyzing}. 
In the same vein, carefully designed GNN models can construct an optimal representations of complex molecular systems that are invariant to translational, rotational, and permutational symmetries. Translation and rotation invariance of can be achieved by introducing input features like radial distances and angles, which themselves remain invariant to both translation and rotation. An example can be found in Ref. \onlinecite{sipka2023constructing}, which proposed a pretrained GNN-based representation generator on translation and rotational invariant features with atom-centered symmetry functions for OP construction.  On the other hand, in order to classify the overall state of a given structure as a given phase of matter, a natural way is to apply pooling aggregators, where pooling over fully connected layers enforces correspondences between mappings of feature space and classes, and this leads to permutation invariance in GNN models.
%This means the output representation is invariant to changing the order of atoms in a graph. 
These properties in combination make GNN models useful in capturing characteristics of the state of highly ordered crystal structures. 

While the above approaches are elegant and powerful, there are not yet many approaches where the learned representations from a GNN are biased to enhance the sampling of nucleation processes. The only notable exception is a very recent preprint. \cite{dietrich2023machine} In this work, we develop a GNN-based autoencoder framework for acquiring low-dimensional representations that is then used in enhanced sampling of crystal nucleation in two different systems. A GNN model in an encoder-decoder setup is trained to precisely classify crystalline structures using local environments. It ensures permutation, translation, and rotation invariance in its latent outputs. A global pooling layer achieves invariance to permutational symmetry, while translation and rotation invariances arise from carefully chosen input features. These compact learnt variables can be easily integrated into various enhanced sampling techniques, thanks to the differentiability of machine learning models. To demonstrate the efficacy of our approach, we apply it to two challenging systems, namely iron and glycine, in the context of nucleation from the melt using metadynamics. Our results show the obtained latent variables capture key configurational features from the training dataset and are able to enhance the samplings as evidenced by frequent back-and-forth transitions.
Our work has complementary aspects to Ref.\onlinecite{dietrich2023machine}. While we benchmark directly on input atomic configurations, allowing the GNN model to probe their structural differences, the model from Ref.\onlinecite{dietrich2023machine} provides an accurate and efficient method for reconstructing conventional order parameters by including their information in the loss function.
We then conduct comprehensive thermodynamic analyses, focusing on the computation of free energies, which exhibit excellent agreements with existing literature regarding the stability rankings of various allotropic and polymorphic structures. 

\section{Methods}
\label{sec:method}
As introduced above, simulating rare events like crystal nucleation in simulations require both associated progress coordinates and enhanced sampling methods to increase movement along these progress coordinates. Here we provide detailed information on both of these aspects in the following three subsections: we first introduce graphs and graph neural nets in Sec.~\ref{sec:ecc}, and we present details on how these graphs are prepared (Sec.~\ref{sec:dataset}). Then, we summarize the sampling method well-tempered metadynamics, in Sec.~\ref{sec:metad}. We provide details on setting up the simulations in the SI and descriptions of the notations in the machine learning model in Tab.\ref{tab:notation}.

\subsection{Graph Neural Networks (GNN) based model}
\label{sec:ecc}
A graph $G=(V,E)$ has two primary components: vertices (or nodes) $V$ and edges $E$. Node (X) and edge (L) embeddings correspond to labels on vertices and edges respectively. In this work, we use a k-nearest neighbor (kNN) algorithm \cite{fix1989discriminatory} to construct the neighborhood $\mathcal{N}_v$ where k is a tunable parameter. In Fig.~\ref{fig:gnn} panel 1, a graph with 5 vertices is presented as an example with $N_X$ number of node features and $N_L$ number of edge embeddings, these feature tensors eventually serve as input to GNN models. For simplicity, node features are set to 1 for systems studied in this work and edge features are selected to be translation and rotation invariant (which we discuss in the next subsection, Sec.\ref{sec:dataset}). Models designed in this work adopt a supervised learning scheme which therefore requires ground-truth labels of different crystal phases for computing and minimizing the loss function. The atom-wise labels are generated with different baseline classifying methods and the graph-level label is determined by the leading population of fingerprints on nodes. 
%The atom-wise labels are generated by adaptive Common Neighbor Analysis (aCNA) \cite{honeycutt1987molecular} and the graph-level label is determined by the leading population of aCNA fingerprints on nodes. 

\begin{table}[b]
    \centering
    \caption{Notations for the machine learning model used in this paper}
    \begin{tabular}{p{2.5cm}  p{5.5cm} }
    \hline  
        Notations & Descriptions \\ 
    \hline
         $V$, $X$ & Nodes and node embeddings  \\
         $E$, $L$ & Edges and edge embeddings \\
         $N$ & Number of vertices (nodes) \\
          $\mathcal{N}_v$ & Neighborhood set of node $v$ \\
         $l$ & Model layer index \\
         $F$ & Edge convolution network in ECC layers \\
         $y$, $Y$ & Global and local predictions \\
         $\tilde{y}$, $\tilde{Y}$ & Graph and node labels (one-hot) \\
         $Z$ & Node latent variables \\
         $z$ & Global (pooled) latent variables \\
         $\mathcal{L}$ & Loss function \\ 
         $\beta$ & Hyperparameter in loss function \\ 
         %$\textbf{x}$ & Particle position tensor \\ 
         %$r_{ij}$ & Euclidean distance between $i$ and $j$ \\ 
         %$\Delta G$ & Free energy difference \\
    \hline 
    \end{tabular}
    \label{tab:notation}
\end{table}
As shown in Fig.~\ref{fig:gnn}, the learning scheme is composed of three parts: creating graph data, training GNN models to learn low-dimensional order parameters (OPs), and performing enhanced sampling along a further reduced space. In this work this is done through an embedded autoencoder framework that allows one to obtain a low-dimensional representation for generic enhanced sampling methods. Specifically, starting from selected node and edge features (described separately in the next subsection) as input features, the encoder, colored in green in Fig.~\ref{fig:gnn}, takes the input and compresses them into a relatively lower-dimensional local descriptor $Z$. For the applications shown in this work, $Z$ is $(N,2)$ dimensional, where $N$ is the total number of nodes in a graph. The local descriptor (in blue in Fig.~\ref{fig:gnn}) predicts the structure of individual nodes, which could be atoms or molecules, given the information about their neighborhood. In addition, a global decoder, shown in yellow, classifies the entire input graph by coupling it with a global pooling layer. Here we choose to use the global mean pooling layer which avoids the effect of system size, compared to other schemes such as sum or max poolings. In other words, this makes the trained model size-agnostic (i.e., transferable to systems of the same species of any size). The output of the global mean pooling operation, $z$, is generally two-dimensional and is biased in enhanced sampling. 
%\pt{consistent in figures/text with $Z$ or $z$}

The graph convolutional layers, specifically edge-conditioned convolution (ECC) layers \cite{simonovsky2017dynamic,gilmer2017neural,moradzadeh2023topology}, allow message passing of node and edge features of the linked neighboring nodes $\mathcal{N}_v$ into the individual vertex through convolution operations, as described in Eq.\ref{eq:message}:
\begin{align}
X^l(i) = \frac{1}{|\mathcal{N}_v|}\sum_{j\in{\mathcal{N}_v}} F^l(L(j,i);w^l)X^{l-1}(j)+b^l, 
  \label{eq:message}
\end{align}
where $l$ is the layer index in the neural network, $w$ and $b$ are learnable weights and biases of the network. A filter network \cite{jia2016dynamic} $F^l$  parameterized by weights $w$ outputs an edge-specific weight matrix given edge attributes $L(j,i)$. We keep the same machine learning architecture for the different systems studied in this work. In particular, we keep zero hidden layers in both decoders to maximally optimize the ability of the encoder to classify different crystal structures. 

The learning objective $\mathcal{L}$ of this model consists of a sum of two cross-entropy losses provided in Eq.~\ref{eq:loss}, where classes in this work correspond to crystal structures.
The first term, which is a local prediction term, computes the cross-entropy loss of the node logits $Y_{i,c}$ of node $i$ class $c$ with respect to the node-level ground truth $\tilde{Y}_{i,c}$ which is intrinsically a binary indicator (0 if the node does not belong to class $c$ and 1 it does so) and then sums over all classes and nodes.  
%\pt{be careful and precise with use of symbols. every symbol should be defined. How does $\textbf{Y}$ relate to other things so far? Y without textbf shows up in fig 1 but is not defined there. are Y and $\textbf{Y}$ same? if so use same symbols. y from next term also shows up in fig 1 but is not defined anywhere. in fig 1, y is a vector with 4 entries. explain explicitly what that means. to simplify thing, I recommend putting a table after figure 1 which summarizes all symbols, y, Y, $\textbf{Y}$, z,Z yhat etc. and be careful with symbols changing bold to non-bold or caps to small!} 
The second term, which is a global prediction term, calculates the cross entropy between the $c^{th}$ class graph prediction ($y_c$) and the target ($\tilde{y_c}$). A hyperparameter $\beta$ is introduced to control the relative importance in local and global prediction. 
%From our numerical investigations, the model is fairly robust to this parameter, and the values are typically ranged in $[0.1,1]$. 
%\pt{define classes, mention explicitly something like "here the index $i$ sums over ..."}

\begin{align}
\mathcal{L}=-\sum_{i}^N\sum_{c\in{classes}} \tilde{Y}_{i,c}\log{Y_{i,c}} - \beta\sum_{c\in{classes}} \tilde{y}_{c}\log{y_{c}}.
\label{eq:loss}
\end{align}

\begin{figure}
  \centering
  \includegraphics[width=8cm]{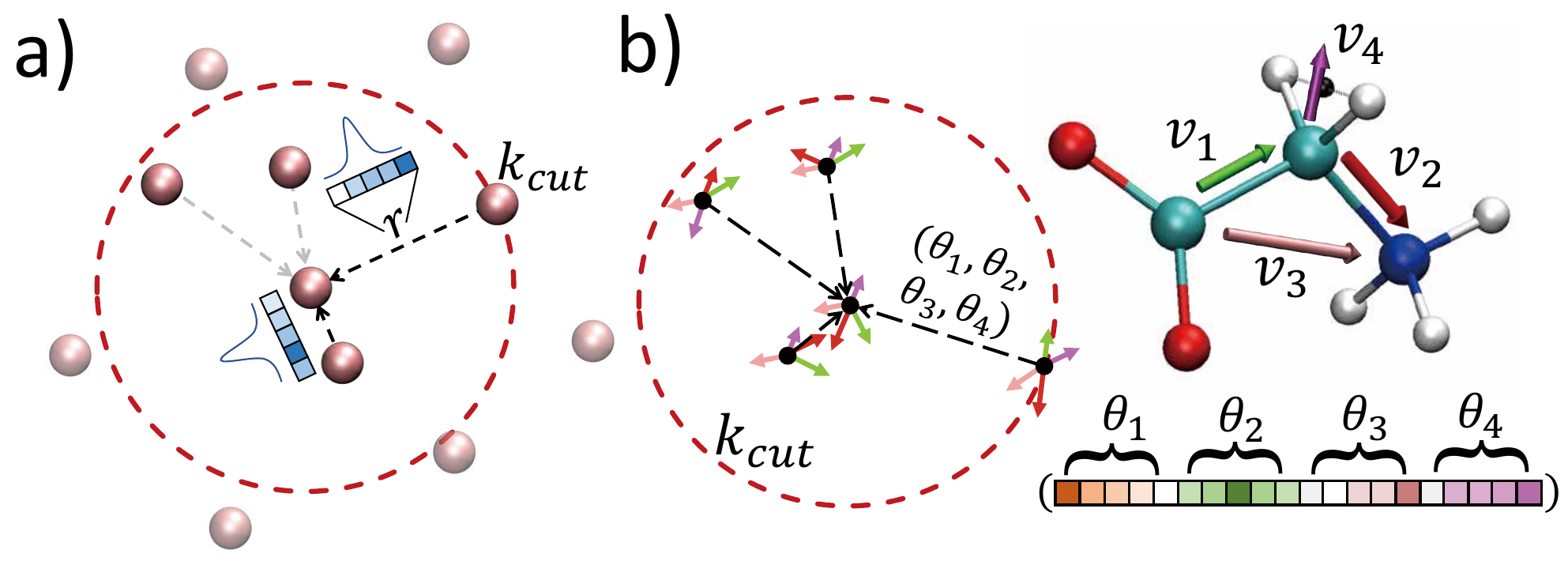}
  \caption
  {Geometric representations of (a) iron particles and (b) glycine molecules. Gaussian basis functions are applied to featurize edge attributes individually and node features are set to be one for both systems. The neighborhood is defined with the k-nearest neighboring algorithm ($k_{cut}=$4 as an example). $v_i$'s are intramolecular characteristic vectors and $\theta_i$'s are the corresponding intermolecular angles formed by specific $v_i$. $v_1$ is the C-C$_\alpha$ vector, $v_2$ is the N-C$_\alpha$ vector, $v_3$ is the C-N vector and $v_4$ is the C$_\alpha$-H$_{center}$ vector. Atoms are colored with respect to their species: irons in mauve, oxygens in red, carbons in cyan, nitrogens in blue, and hydrogens in white. Ghost particles (i.e., center of mass of glycine and center of hydrogen atoms) are in black. 
  }
  \label{fig:neighbor}
\end{figure}

\subsection{Dataset Preparation}
\label{sec:dataset}
Like all data-driven methods, prior information needs to be provided to train our model. In this work, the model uses all-atom coordinates from MD simulations initiated from different perfect crystal structures. The training data for iron allotropes was generated by LAMMPS built-in lattice functional; while the molten phase was prepared by random insertion of iron particles. Separately, four supercells composed of 432 $\alpha$-Fe, 256 $\gamma$-Fe, 180 $\epsilon$-Fe, and 285 molten Fe were constructed and 1 $ns$ short MD trajectories were initiated accordingly. These structures were equilibrated at respective temperatures where they are expected to be stable. Specifically, this was  1000 K for $\alpha$-Fe, 900 K for $\gamma$-Fe, 900 K for $\epsilon$-Fe, and 2000 K for molten Fe. A total of 2000 frames of MD snapshots, corresponding to 500 frames for each configuration, were then converted into graph representation and trained via the proposed model discussed above. The node feature was set to be unity. A sparse adjacency matrix 
%\pt{be consistent with was or is, were or are. I prefer past tense} 
was constructed on k-nearest neighbors (k-NN) metric. \cite{eppstein1997nearest}
%, which is equivalent to the radius neighbor algorithm \cz{(no citation for this, I discussed with Luke months ago, and we both think the two are interchangeable here. If the scenario changes solvated system, it will make a difference)} with appropriate parameterizations. 
Each pair of linked nodes was attributed the radial distance as the edge feature, followed by a Gaussian basis function introduced in Schnet shown in Fig.\ref{fig:neighbor}a) .\cite{schutt2017schnet} An edge feature, $L$, can be obtained by expanding distance, $r$, into $t$ slices as follows:
\begin{align}
L_t(\textbf{x}_i-\textbf{x}_j)=\exp(-\gamma ( r_{ij} - \mu_{t} )^2 ) ,
\label{eq:basisfxn}
\end{align}
where $\bf{x_i}$ is the position tensor for node $i$, and $r_{ij}=||\textbf{x}_i-\textbf{x}_j||$ guarantees translation and rotation invariance of the GNN model. 

Crystal structures of glycine polymorphs were obtained from the Crystallography Open Database. \cite{Grazulis2009COD} In a similar manner as for iron, different supercells consisting of 128 molecules as $\alpha$-glycine\cite{dawson2005alphagamma-gly}, 128 as $\beta$-glycine\cite{iitaka1960beta-gly}, 108  as $\gamma$-glycine\cite{dawson2005alphagamma-gly}, and 137 as liquid glycine (denoted as $\ell$-glycine) were built using unit cells of glycine polymorphs prepared with the Mercury package. \cite{macrae2020mercury} MD simulations of all glycine polymorphs were carried out at 300 $K$; while the liquid glycine was obtained at around its melting temperature of 500 $K$. All MD simulations are of length 2 $ns$ and 2000 frames were obtained from each simulation for training the model. Individual glycine molecule is treated as one entity with the corresponding center of mass (i.e., one node per molecule; black dots in Fig.\ref{fig:neighbor}b)) whose feature is 1. The leading 6 nearest neighbors are defined as neighboring molecules referring to the position of the center of mass of each molecule. Different from iron, edge attributes are concatenations of four intermolecular angles under Gaussian basis function into one long feature vector as shown in Fig.\ref{fig:neighbor}b) bottom right. The angles are defined by the intramolecular vectors: $\nu_1$ is the C-C$_\alpha$ vector, $\nu_2$ is the N-C$_\alpha$ vector, $\nu_3$ is the C-N vector and $\nu_4$ is the vector of C$_\alpha$ and the geometric center of the two associated hydrogens (see Fig.\ref{fig:neighbor}b) for illustrations). %\sout{We deliberately removed radial distance from the edge feature list to avoid possible problems of direct convolution of features in different units. } 
%The Gaussian centers to distances ranged in $0nm \leq \mu_t \leq 0.8nm$ every 0.08 $nm$ with $\gamma=156.25\ nm^{-2}$. 

\subsection{Well-Tempered Metadynamics}
 \label{sec:metad}
 The low-dimensional latent variables learnt through the global mean pooling operation (Fig.~\ref{fig:gnn}) serve as low-dimensional descriptors of various competing phases. Due to the loss function in Eq.~\ref{eq:loss} these capture both local and global information, making them well-suited for driving short-range and long-range fluctuations relevant to nucleation. Here we do so by performing well-tempered metadynamics  (WTmetaD)\cite{valsson2016enhancing} along these variables, while expecting that our protocol should be fully amenable to other enhanced sampling approaches.
%It is insufficient to evaluate the ability of the learnt low-dimensional latent representation by simply monitoring loss function decreases or overlapping of different classes in scatter plots (left panel in Fig.~\ref{fig:gnn}3). A variable that classifies different metastable states is not equivalent to an optimal reaction coordinate. However, this feature is often considered a prerequisite (which should be visualized before further examinations). We, therefore, performed an enhanced sampling method, namely, well-tempered metadynamics (WTmetaD) \cite{valsson2016enhancing} along these learnt latent representations to examine whether state-to-state back-and-forth transitions are visited when biasing.

In WTmetaD, history-dependent Gaussians are deposited along pre-defined biasing variables reaction coordinates to facilitate state-to-state back-and-forth transitions between different metastable states the system would normally be trapped in. We refer to Ref.\onlinecite{valsson2016enhancing} for further details of WTmetaD. Here we used the latent variables $(z_1,z_2)$ from Fig.~\ref{fig:gnn} as the variables being biased. Other parameters used in performing WTmetaD simulations are reported in Tab.~\ref{tab:metad_parameter}. Iron nucleation simulations were performed with LAMMPS-23Jun2022 simulator \cite{LAMMPS} and glycine simulations were performed with GROMACS-2021.6 MD engine.\cite{berendsen1995gromacs} Both packages were patched to PLUMED-2.8.1 with the Pytorch module enabled. \cite{plumed2019nature, plumed2} Codes for reproducing the simulations in this work are available at \href{https://github.com/tiwarylab/GNN-PLUMED-Nucleation}{Github}. %\pt{update repo for this paper specifically}

\begin{table}[b]
    \centering
    \caption{WTmetaD Parameters}
    \begin{tabular}{c|c|c|c|c|c|c}
    \hline \hline 
        System & $\omega$($k_bT$) & $\gamma$  & $\sigma_1$ (RC unit) & $\sigma_2$ (RC unit) & $T$ ($K$) & pace \\ 
          \hline 
         Iron & 1.0 & 50 & 0.2 & 0.2 & 1800 & 500  \\
         \hline
         Glycine & 2.0 & 100 & 0.15 & 0.1 & 500 & 500 \\ 
         
    \hline \hline
    \end{tabular}
    \label{tab:metad_parameter}
\end{table}

%\subsection{Simulation Setup \pt{move this section fully to SI, while updating intro of sec II to reflect this change}}
% \label{sec:setup}

\section{Results and Discussions}
\label{sec:results}
We evaluate the ability of the GNN-learnt low-dimensional latent representations to enhance sampling by performing well-tempered metadynamics for the two selected representative systems, namely, iron (Sec.\ref{sec:iron}) and glycine (Sec.\ref{sec:glycine}) initiated from their molten or liquid phases. Iron as one of the most abundant elements on Earth has received significant interest given its importance in steels and alloys and in geology. Many allotropes of pure iron exist, which are the body centered cubic (BCC) $\alpha$-Fe, the face centered cubic (FCC) $\gamma$-Fe, the hexagonal close packing (HCP) $\epsilon$-Fe and the BCC $\delta$-Fe. In addition to iron, we assess the reliability of our protocol on the nucleation of polymorphs of the simplest amino acid glycine. This is an important system as physical properties in different glycine polymorphs vary which can change the effect of glycine as an inhibitory neurotransmitter. \cite{Lynch2004} Furthermore, the existence of many possible space groups complicates the problem in molecular crystals in general.\cite{boldyreva2003polymorphism} Three polymorphs, namely $\alpha$-,$\beta$-, and $\gamma$-glycine, exist in zwitterionic glycines at ambient conditions. In particular, $\alpha$-glycine (space group: P2$_1$/n) and $\beta$-glycine (P2$_1$) are monoclinic and $\gamma$-glycine (P3$_1$) is trigonal.  We compute free energy differences between these different metastable allotropes/polymorphs and compare with respective literature.

\begin{figure*}
  \centering
  \includegraphics[keepaspectratio, width=16cm]{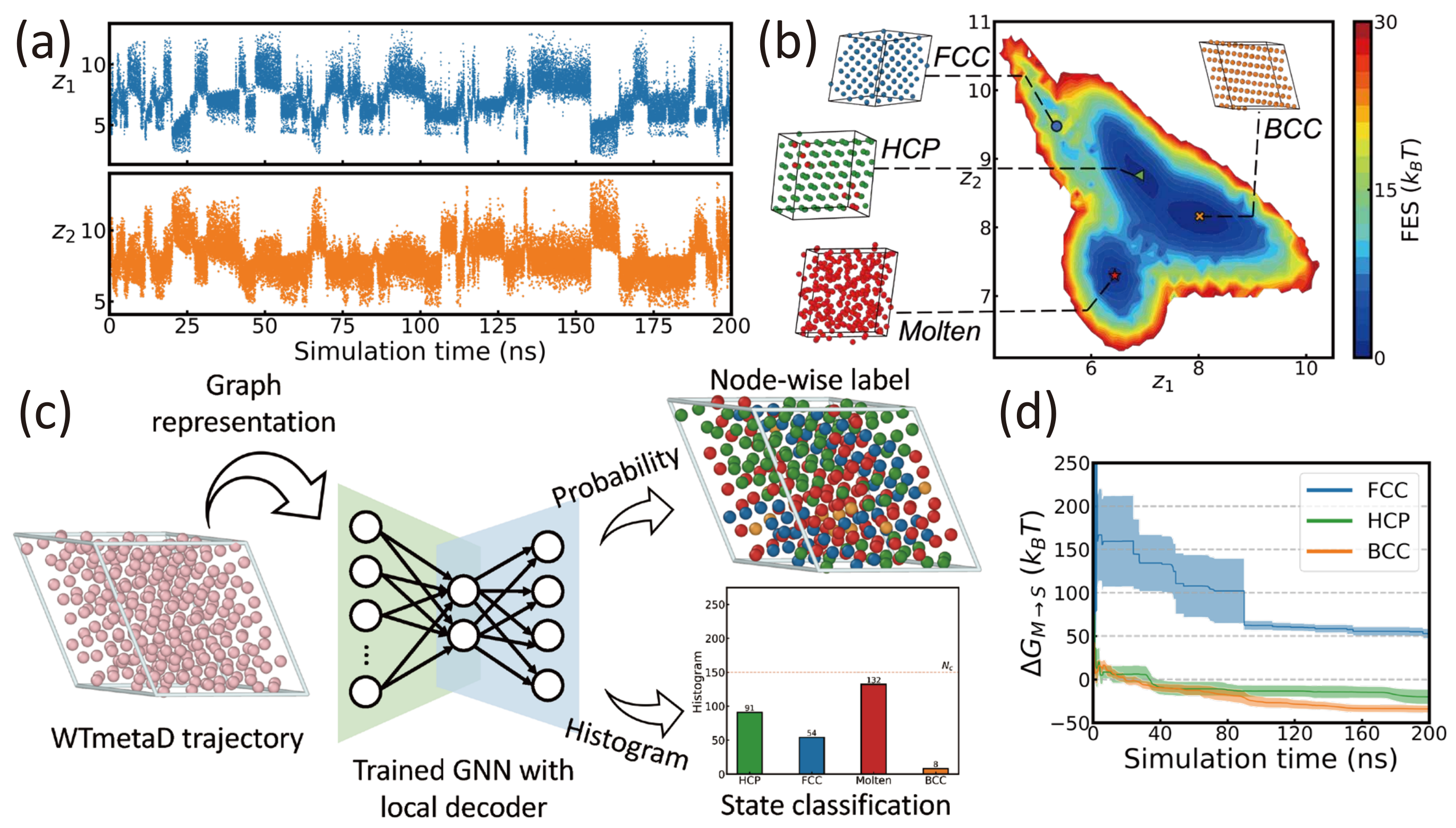}
  \caption
  {Results of WTmetaD simulations of iron nucleation from its melt. Machine learnt latent representations $z_1,z_2$ as a function of simulation time in panel a) shows frequent state-to-state transitions. Subplot b) 
 shows reweighted FES (free energy surface) in the latent variable space ($z_1,z_2$) with landmarks of sampled configurational snapshots from OVITO package. \cite{Stukowski2010ovito} Iron particles are color-coded with respect to the adaptive common neighbor analysis method. \cite{Stukowski2012} Subplot c) shows the workflow of post-processing the obtained WTmetaD trajectories by passing them through the full GNN model with the local decoder. The state labels are generated from the output prediction and consequently applied for computing the time series of free energy difference of solid states of interest (S) with respect to the molten phase M in subplot d). Computed standard errors are represented as shaded regions. 
  }
  \label{fig:fe_metad}
\end{figure*}

\subsection{Phase Transitions in Pure Iron}
\label{sec:iron}
%move earlier Iron as one of the most abundant elements on Earth receives many research interests given its importance in steels and alloys. Many allotropes of pure iron exist, which are the body centered cubic (BCC) $\alpha$-Fe, the face centered cubic (FCC) $\gamma$-Fe, the hexagonal close packing (HCP) $\epsilon$-Fe and the BCC $\delta$-Fe.
Experimental measures suggest $\alpha$-Fe remains stable at temperature less than 1184 $K$; while $\gamma$-Fe is stable between 1184 $K$ to 1665 $K$. When above 1665 $K$, $\gamma$-Fe transforms to another BCC structure, $\delta$-Fe until sublimation. In addition, the HCP $\epsilon$-Fe is stable at a pressure greater than 13 $GPa$. \cite{lee2012atomistic,ou2017molecular} The enrichment of crystal packings makes iron a challenging system and an excellent test case for our proposed protocol. 

The graph data for iron particles is shown in Fig.\ref{fig:neighbor}a). Since the system of interest here is in single species, the feature of individual nodes is set to be 1. We set a k-NN cutoff ($k=50$) that is much larger than the coordination number (number of neighboring particles in the first shell) of close packing structures. This helps gain information from nodes several hops away in order to account for the structural similarities in FCC (abcabc...) and HCP (ababab...) crystals. %\sout{(remove this if you are not using it anywhere)An alternative way is to use multiple convolutional operations for message passing.}
The edge features are the Gaussian expanded radial distance as introduced in Methods (Sec.~\ref{sec:dataset}). The edge feature is an expanded radius distance in 10 dimensions.

We find that well-tempered metadynamics biasing along the 2-dimensional latent variable learnt from GNN leads to robust, multiple state-to-state transitions without any hysteresis, as can be seen from the time series in Fig.\ref{fig:fe_metad}a). The associated free energy surface is then constructed following the appropriate reweighting scheme. \cite{tiwary2015jpcb} Three distinct energy basins correspond to the initial molten state, HCP iron, and BCC iron (Fig.\ref{fig:fe_metad}b)). On the contrary, no distinct basin is sampled for the FCC state of iron, which suggests such a configuration is thermodynamically less stable at the simulation temperature of 1800 $K$. 
This is expected because FCC is reported to be the least stable allotrope among other forms from both experiments \cite{guillermet1984assessment} and zero temperature calculations \cite{etesami2018molecular,lee2000second}. 
%Additionally, benefit from the robustness of the WTmetaD method, crystal structures that are not included in the training set are visited and the structure of those are determined to be simple cubic (see SI for snapshots). This shows WTmetaD can largely enhance the sampling even under suboptimal RC when certain metastable states are not guaranteed to be distinguished by the proposed RC.

\begin{table}
    \begin{center}
    \caption{Free energy differences between melt/liquid and crystal structures sampled in WTmetaD simulation with respect to the initial molten (M)/liquid ($\ell$) phases.}
    \begin{tabular}{c|c|c}
    \hline \hline
        System & Transition & Free energy difference ($kJ/mol$)  \\ 
         \hline \hline
         \multirow{4}{*}[0.6em]{Iron} & M$\rightarrow$FCC & 52.63 $\pm$ 5.63 \\
         & M$\rightarrow$HCP & -20.15 $\pm$ 8.52  \\
         & M$\rightarrow$BCC & -34.06 $\pm$ 4.20 \\\cline{2-3}
         
         \hline
         \multirow{4}{*}[0.6em]{Glycine} & $\ell\rightarrow\alpha$ & 56.43 $\pm$ 16.92  \\
         & $\ell\rightarrow\beta$ & 658.87 $\pm$ 74.72  \\
         & $\ell\rightarrow\gamma$ & 234.11 $\pm$ 35.41  \\\cline{2-3}

         \hline 
    \end{tabular}
    \centering
    %\centering  \parbox{7.25cm}{$^*$ Solid solid transitions are counted separately, e.g. a transition of urea or glycine solution (L) $\rightarrow$ solid form 1 $\rightarrow$ solid form 2 is  counted as L $\rightarrow$ S1 and L $\rightarrow$ S2.}
    \label{tab:fe_diff}
    \end{center}
    
\end{table}

\begin{figure*}
  \centering
  \includegraphics[keepaspectratio, width=16cm]{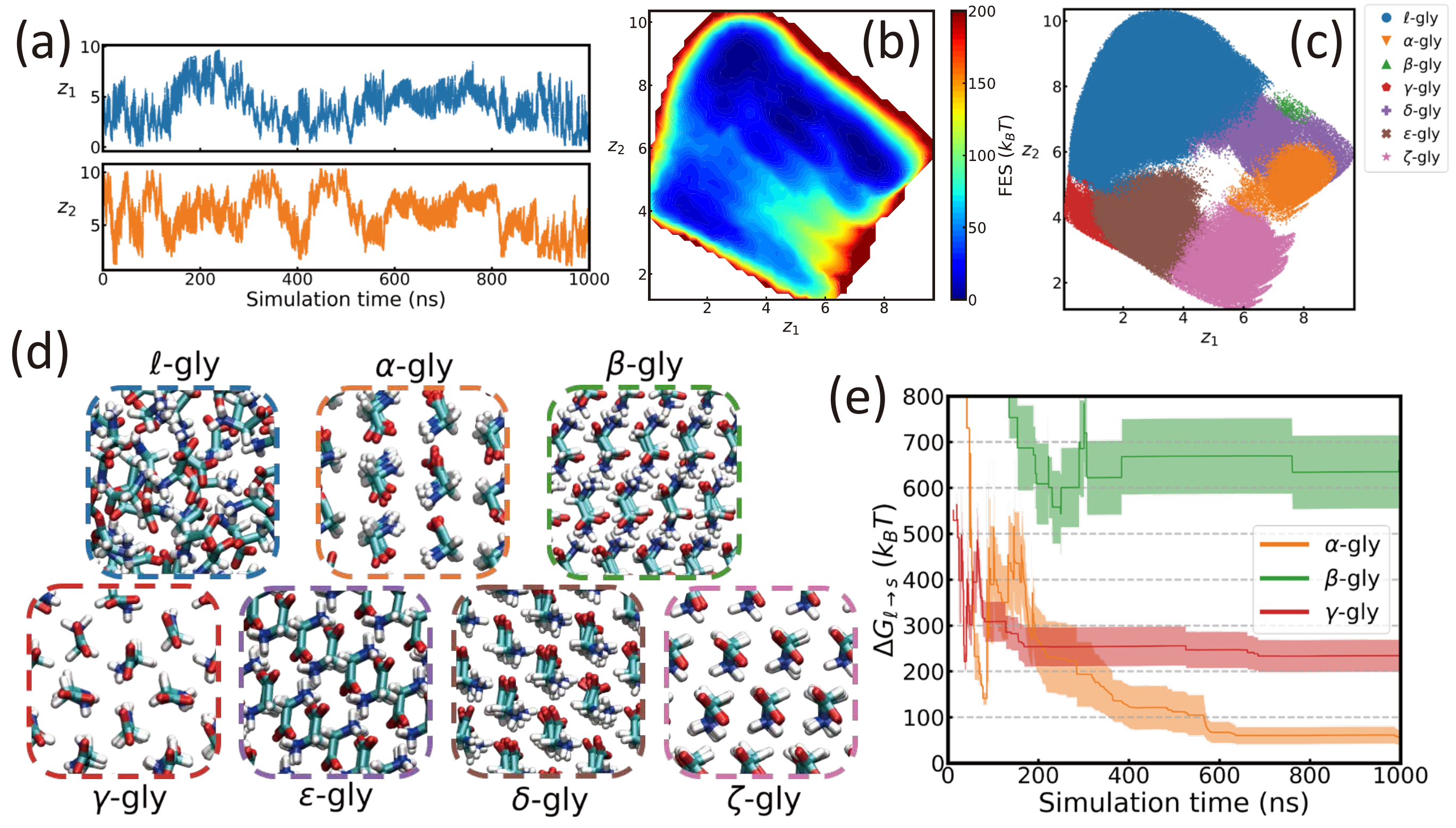}
  \caption
  {Results of WTmetaD simulations of the nucleation of glycine zwitterions. a) shows machine learnt latent representations $z_1,z_2$ vs. simulation time. Subplot b) 
 shows reweighted FES (free energy surface) in the latent variable space ($z_1,z_2$) with multiple basins observed. Subplot c) is the scatter plot of liquid and solid glycines in ($z_1,z_2$) space classified by post-trained GNN model, and d) consists of snapshots to each corresponding class rendered from Visual Molecular Dynamics (VMD). \cite{HUMPHREY1996VMD} Free energy difference between states of interest vs. simulation time in e) shows convergence of the production runs. Standard errors are represented as shaded regions. 
  }
  \label{fig:gly_metad}
\end{figure*}

An advantage of the proposed ML model is that its local decoder (Fig.\ref{fig:gnn}) provides an estimation, in a probabilistic sense, of the class or crystalline structure of individual nodes belonging to. \cite{steinhardt1983op, honeycutt1987molecular, faken1994systematic} The populations of different crystal packings can then be computed by the summation of individual weights from the output $Y$ of the local decoder, $N_{class}=\sum_{i}^{N} Y_{i,class}$ as shown schematically in Fig.\ref{fig:fe_metad}c). Additionally, we present results in the form of a 2-dimensional free energy surface in the SI for indicators of potential solid-solid transitions. The free energy difference $\Delta G$ can be computed between phases and, as an example, the equation between liquid, $\ell$, and solid, $s$, is shown as follows:
\begin{align}
\Delta G_{\ell\rightarrow s} = -k_BT \ln \left (\frac{\langle H(N_s-N_c)\rangle}{\langle H(N_\ell-N_c)\rangle} \right) ,
\label{eq:fediff}
\end{align}
where $\langle \cdot \rangle$ is the reweighted unbiased average and $H$ is the Heaviside step function with a size cutoff $N_{c}$. The tabulated free energy differences between metastable crystalline states and the starting molten state are plotted in Fig.\ref{fig:fe_metad}d) and reported in Tab.\ref{tab:fe_diff} with a threshold value of 150 in Eq.\ref{eq:fediff}, which means frames are categorized into the corresponding state for free energy computation when $N>150$ (within a total of 285 Fe irons in the box). In addition, Fig.\ref{fig:fe_metad}d) suggests that the WTmetaD simulations performed are well-converged. The values were averaged over four independent production runs of 200 $ns$. The thermodynamically most stable configuration of iron is its BCC form with a free energy difference $-25\pm4\ kJ/mol$ relative to molten iron. The least stable FCC iron has a free energy difference of $65\pm3\ kJ/mol$ relative to molten iron. However, owing to strong finite size effects, it is more meaningful to draw qualitative comparisons to the experimental and computational works of literature. As reported in Ref.\onlinecite{guillermet1984assessment}, the experimental measured free energy difference between BCC and FCC iron, $\Delta G_{BCC-FCC}$, is $6.66\ kJ/mol$ and $\Delta G_{FCC-HCP}$ is $-2.22\ kJ/mol$ at room temperature. Zero temperature calculations \cite{lee2000second,etesami2018molecular} also show similar measures which are in good agreement with the stability rankings obtained in this work, BCC $>$ HCP $>$ FCC iron in decreasing order.

\subsection{Nucleation of Glycine from Melt}
\label{sec:glycine}
The developed model is then assessed to a more complicated molecular system, glycine. Geometric data for zwitterionic glycine is slightly more intricate than that of iron (see \hyperref[sec:dataset]{Methods} for detailed information). As more degrees of freedom are incorporated into the graphs of glycine molecules, a smaller neighborhood of individual molecules is defined with only 6 closest neighbors ($k=6$). Intermolecular angles of characteristic vectors, $\nu_1$, $\nu_2$, $\nu_3$, and $\nu_4$ are again expanded under basis functions which leads to 40-dimensional edge features between linked nodes.

Fig.~\ref{fig:gly_metad} summarizes results from WTmetaD simulations biasing latent representations $z_1,z_2$ learnt by the GNN model. Several transitions can be identified by evaluating the time series in subplot a) along with multiple distinct wells in the reweighted free energy surface of ($z_1,z_2$) space (Fig.~\ref{fig:gly_metad}b)). However, after closely examining the obtained trajectories with visualization tools, we found that more polymorphic glycines were sampled even though the model was trained only on solid glycine in its three well-studied ambient products synthesized experimentally. These new polymorphs that our simulations visit have been however reported previously in Ref.\onlinecite{dawson2005alphagamma-gly,Bull2017} as high-pressure structures. We thereby trained a new GNN model as a classifier on liquid glycine and all associated crystals including structures found under the effect of different pressures, and this leads to in total of six polymorphs namely $\alpha$-, $\beta$-, $\gamma$-, $\delta$-, $\epsilon$-, and $\zeta$-glycine. The notations for the solid glycines are adopted from Ref.\onlinecite{Bull2017}. Here, we only briefly describe the three additional glycine polymorphs since they are not the main focus of this work and their relative stabilities remain unclear: $\alpha$-glycine remains stable to pressures up to 23 $Gpa$, meanwhile, $\beta$-glycine undergoes phase transition to $\delta$-glycine ($P2_1/c$) at 0.8 $GPa$, $\gamma$-glycine transforms to $\epsilon$-glycine ($Pn$) under application of pressure and the process complete at between 4 to 5 $GPa$, and decompression of $\epsilon$-glycine leads to $\zeta$-glycine ($I1$).\cite{Bull2017,boldyreva2021glycine}

The distributions of individual glycine polymorphs are shown as scatter points in ($z_1,z_2$) space in Fig.~\ref{fig:gly_metad}c) with cutoff value $N_c=65$, and the corresponding snapshots are provided in Fig.~\ref{fig:gly_metad}d). All observations above suggest the trained GNN latent representations capture the configurational information among various polymorphic structures of glycine molecules. Structures which are close to each other in configuration space (i.e. the associated high-pressure components) are sampled at ease with robust WTmetaD simulations.

The free energy difference $\Delta G_{\ell\rightarrow s}$ as a function of simulation time shown in Fig.~\ref{fig:gly_metad}e) is computed from Eq.~\ref{eq:fediff} with a value of cutoff 65 which is set slightly larger than half of the population of glycine in the simulation cell. The figure is averaged over 9 independent runs of 1000 $ns$ (see the SI for full $\Delta G_{\ell\rightarrow s}$ vs. simulation time plot) and the exact values are reported in Tab.~\ref{tab:fe_diff}. This indicates the thermodynamic stability of ambient glycine poylmorphs ranks in $\alpha$-gly $>$ $\gamma$-gly $>$ $\beta$-gly at 500 $K$ which is in consistent with the stability ranking reported in the literature which $\gamma$-gly is the most thermodynamically stable at ambient temperature and the densest $\alpha$-gly becomes the most stable at temperatures above 440 $K$. \cite{boldyreva2021glycine} 
In addition, we also identified a transition pathway of $\gamma$-gly to $\epsilon$-gly to $\zeta$-gly, while no direct transitions from $\gamma$-gly to $\zeta$-gly, supported by 2-dimensional free energy analyses (see SI for details), which was also observed and reported experimentally. \cite{goryainov2006raman} 
Overall, this shows that %\pt{rephrase this part of the sentence? even with GNN architecture in simple convolution operations and appropriate feature selection}, 
the trained model is robust in learning structural properties for classifying configurations with simple graph convolution operations and readily computable features. Benefiting from the WTmetaD method, introducing biases along these low-dimensional latent representations validates the feasibility of obtaining relative free energy differences between competing allotropes/polymorphs starting only from their chemical identity and possible target structures.

\section{Conclusion}
\label{sec:conclusion}
Computational methods for investigating crystal nucleation have recently shown their strength in providing high temporal- and spatial-resolution descriptions. \cite{sosso2016crystal, blow2021seven} Due to the timescale limitations resulting from the rare event nature of nucleation, it is however necessary to perform enhanced sampling molecular dynamics. Most enhanced sampling methods involve biasing or following selected low-dimensional descriptors, and methods of constructing these descriptors remains to be an active field of research. Recent advancements in machine learning, particularly graph neural networks, have made it possible to achieve a better understanding of crystal nucleation from a perspective of learning relative slow modes \cite{liu2023graphvampnets}, or on efficient computation of order parameters \cite{dietrich2023machine}. 

In this work, we have introduced a data-driven GNN-based representation learning model within an autoencoder framework to extract low-dimensional variables from configurational features found in experimental crystal structures. These variables serve as the key inputs for performing enhanced sampling methods. Our method employs straightforward convolutions and pooling techniques. To validate the usefulness of our machine learning pipeline, we studied nucleation of different allotropes and polymorphs of iron and glycine respectively from the melt. We biased the GNN based latent variables in well-tempered metadynamics and were able to achieve multiple back-and-forth state-to-state transitions and converged free energy estimates, both hallmarks of reliable sampling. This proves the robustness and potential of our graph neural network learnt variables for enhanced sampling. The thermodynamic stability rankings among allotropes or polymorphs are in agreement with experimental measures.

The protocol proposed here can be further improved in many ways, and here we highlight some possible avenues for future research. The current network only consists of convolution and pooling operations and it should be possible to introduce advanced manipulations on graph data, such as the attention mechanism, to better interpolate and even extrapolate configurational information of complex species. \cite{banik2023cegann,xie2018crystal} Secondly, an obstacle to drawing quantitative comparisons to literature is due to finite size effets. To address this issue, various methods have been developed, including the introduction of constant chemical potential ensemble \cite{karmakar2023non} and analytical correction under the classical nucleation framework \cite{salvalaglio2015molecular,Salvalaglio2016argon,Hussain2022finite-size}. In future work, we hope to combine data-driven approaches with those from statistical mechanics \cite{wang2022data} to not just automate biasing variable discovery, as was done in this work, but also address finite size effects.

%Additionally, it has been shown that learning relative slow motions is necessary for the study of rare events which requires incorporating temporal information.   \cite{Noe2013Tica,pande2017TicaMeta,mardt2018vampnets,wang2019past,Wang2021SPIB,liu2023graphvampnets,ray2023deep,sasmal2023reaction,sun2022multitask} an initial exploration of integrating graph machine learning models into enhanced sampling methods. To validate such a procedure is indeed applicable, we examined it in two representative systems which yield good agreements on the thermodynamic rankings with experimental observations.

%\section{Associated Content}
%\label{sec:associatedcontent}

%\subsection{Data Availability Statement}
%The codes used for reproducing the simulations are available from github.

%\subsection{Supporting Information}
%The Supporting Information is available free of charge at XXX.

%Supporting Information Available: Details on data parameterization and graph neural network model training; Simulations settings; Two-dimensional free energy profiles on iron crystal size (Figure S1); Local decoder on the identification of crystal irons compared to benchmark method (Figure S2 and Table S1); Free energy difference of all glycine polymorphs with respect to liquid glycine as a function of simulation time (Figure S3) and tabulated free energy (Table S2); Two-dimensional free energy profiles on glycine polymorph size (Figure S4).

%\subsection*{Notes}
%The authors declare no competingfinancial interest.

\section*{Acknowledgments}
\label{sec:acknowledgements}
Z.Z. thanks Prof. Mark E. Tuckerman, Prof. Omar Valsson, and Dr. Pablo M. Piaggi for useful advice. Z.Z. also thanks Dr. Zachary Smith and Dedi Wang for help in graph neural nets and C$++$. Z.Z. and P.T. thank Dr. Eric Beyerle, Dr. Bodhi P Vani, Dr. Yihang Wang, Dr. Sun-Ting Tsai, Dr. Luke Evans and Akashnathan Aranganathan for fruitful discussions and Dr. Eric Beyerle, Dedi Wang for proofreading the manuscript. This research was entirely supported by the U.S. Department of Energy, Office of Science, Basic Energy Sciences, CPIMS Program, under Award DE-SC0021009. We are also grateful to NSF ACCESS Bridges2 (project CHE180053) and University of Maryland Zaratan High-Performance Computing cluster for enabling the work performed in this work. 

\bibliographystyle{achemso}
\bibliography{references}

% \clearpage
% \begin{figure*}
  % \centering
  % \includegraphics[keepaspectratio]{toc-final.png}
  % \caption
  % {TOC graphic 
  % }
  % \label{fig:toc}
% \end{figure*}

\end{document}